\documentclass[prl,preprint]{revtex4} 
\usepackage{graphics} 
\usepackage{amsmath} 
\usepackage{graphicx,color}

\newcommand \be  {\begin{equation}} 
\newcommand \beq {\begin{equation}} 
\newcommand \bea {\begin{eqnarray} } 
\newcommand \ee  {\end{equation}} 
\newcommand \eeq {\end{equation}} 
\newcommand \eea {\end{eqnarray}} 
\newcommand{\beqa}{\begin{eqnarray}} 
\newcommand{\eeqa}{\end{eqnarray}}

\newcommand{\rmd}{\text{d}} 

\newcommand{\bol}[1]{{\boldsymbol{#1}}}

\newcommand{\rme}{\text{e}} 
\newcommand{\rmi}{\text{i}}

\begin{document}

\title{Regulatory control and the costs and benefits of biochemical noise} 
\author{Sorin T\u{a}nase-Nicola and Pieter Rein ten Wolde} 
\affiliation{ 
FOM Institute AMOLF, Kruislaan 407, SJ1098 Amsterdam, The Netherlands } 
\date{\today}

\begin{abstract} 
  Experiments in recent years have vividly demonstrated that gene
  expression can be highly stochastic. How protein concentration
  fluctuations affect the growth rate of a population of cells, is,
  however, a wide open question.  We present a mathematical model that
  makes it possible to quantify the effect of protein concentration
  fluctuations on the growth rate of a population of genetically
  identical cells. The model predicts that the population's growth
  rate depends on how the growth rate of a single cell varies with
  protein concentration, the variance of the protein concentration
  fluctuations, and the correlation time of these fluctuations. 
  The model also predicts that when the average concentration of a
  protein is close to the value that maximizes the growth rate,
  fluctuations in its concentration always reduce the growth
  rate. However, when the average protein concentration deviates
  sufficiently from the optimal level, fluctuations can enhance the
  growth rate of the population, even when the growth rate of a cell
  depends linearly on the protein concentration. The model also shows
  that the ensemble or population average of a quantity, such as the
  average protein expression level or its variance, is in general not
  equal to its time average as obtained from tracing a single cell and
  its descendants. We apply our model to perform a cost-benefit
  analysis of gene regulatory control. Our analysis predicts that the
  optimal expression level of a gene regulatory protein is determined
  by the trade-off between the cost of synthesizing the regulatory
  protein and the benefit of minimizing the fluctuations in the
  expression of its target gene. We discuss possible experiments that
  could test our predictions.
\end{abstract}

\maketitle 

\noindent {\bf Author summary}\\ 
\noindent Living cells use regulatory networks in order to respond to
a changing environment. They use gene regulatory networks, for
example, to adjust the optimal expression levels of metabolic enzymes
in response to changing sugar concentrations. Both the regulatory
networks and metabolic networks of living cells are often highly
stochastic. However, how protein concentration fluctuations affect the
growth rate of a population of cells is largely unknown. We present a
mathematical model that makes it possible to predict how protein
concentration fluctuations affect the population's growth rate. The
model predicts that when the expression level of a protein is close to
the value that maximizes the growth rate, fluctuations will always
reduce the growth rate. However, if the average protein expression
level deviates sufficiently from the optimal one, then fluctuations
can enhance the population's growth rate. The reason is that cells
that happen to grow faster will dominate the population. We also apply
our model to investigate the optimal design of a regulatory
network. Our analysis predicts that this is determined by the
trade-off between the cost of synthesizing the proteins that
constitute the regulatory network, and the benefit of reducing the
fluctuations in the network that it controls.

\newpage 

\noindent{\bf \Large Introduction}  \\
Cells continually have to respond and adapt to a changing 
environment. One important strategy to cope with a fluctuating 
environment is to sense the changes in the environment and respond 
appropriately, for example by switching phenotype or behavior. 
Arguably the most studied and best characterized example is the {\em 
  lac} system, where the LacI repressor measures the concentration of 
lactose and regulates the expression level of the metabolic enzyme 
that is needed to consume lactose. In this strategy of responsive 
switching, it is critical that cells can accurately sense and respond 
to the changes in the environment \cite{Kussell05b}. However, both the 
detection and the response are controlled by biochemical networks, 
which can be highly stochastic 
\cite{Elowitz00,Ozbudak02,Elowitz02,Raser05,Kaern05,Becskei05,Rosenfeld05,Yu06,Cai06,Sigal:2006uq}. One 
might expect that noise is detrimental, since it can drive cells away 
from the optimal response curve---the optimal enzyme concentration as 
a function of the lactose concentration \cite{Dekel05b}. On the other 
hand, both reducing noise and creating a regulatory network that 
allows cells to respond optimally can be energetically costly \cite{Dekel05b}, which 
would tend to reduce the fitness of the organism 
\cite{Zhang05}. In this paper, we present a model that 
makes it possible to quantify the effects of biochemical noise on the 
growth rate of a population of cells that respond via the mechanism of 
responsive switching. We then use this model to perform a cost-benefit 
analysis of gene regulatory control, using cost and benefit functions 
that have been measured experimentally \cite{Dekel05b}. This analysis, 
which complements recent work by Kalisky and coworkers 
\cite{Kalisky:2007fj}, predicts that gene regulatory proteins exhibit 
an optimum expression level, which is determined by the trade-off 
between the cost of synthesizing the regulatory protein and the 
benefit of reducing the fluctuations in its target gene. 

It has long been recognized that organisms in a clonal population can 
exhibit a large variation of phenotypes.  Within highly inbred lines, 
for instance, phenotypic variation can still be detected 
\cite{Falconer1996}. More recently, experiments have vividly 
demonstrated that gene expression in uni- and multicellular organisms 
fluctuates strongly 
\cite{Elowitz00,Ozbudak02,Elowitz02,Raser05,Kaern05,Becskei05,Rosenfeld05,Yu06,Cai06,Sigal:2006uq}. The 
fact that fluctuations are not selected out, suggests that the optimal 
fitness requires a certain amount of biochemical noise. However, how 
the growth rate of a population depends upon biochemical noise is 
still poorly understood.  In a constant environment, stabilizing 
selection favors a genotype that leads to a narrow phenotype 
distribution centered around the optimal phenotype in that environment 
\cite{Gavrilets94,Zhang05}. However, cells do not live in a constant 
environment, but rather in one that fluctuates. While one strategy to 
cope with environmental fluctuations is to detect and respond to them 
(responsive switching), an alternative one is to create diversity in 
the population. This can be achieved via the mechanism of stochastic 
switching \cite{Segerbook,Thattai04,Wolf05,Kussell05a}, whereby 
members of the population randomly flip between different phenotypes 
due to biochemical noise. This strategy is particularly efficient when 
the time scales of the environmental fluctuations are either very 
long, such that the investments of constructing an energetically 
expensive response machinery do not pay off \cite{Kussell05a}, or very 
short, i.e. shorter than the time it takes for the population to 
respond to them \cite{Thattai04,Wolf05}. Many examples of this 
strategy exist in nature \cite{vanderWoude04, Balaban04}, and this 
strategy has recently been studied in much theoretical detail 
\cite{Segerbook,Thattai04,Wolf05,Kussell05a}. However, the dominant 
strategy for coping with changes in pH, temperature, the food supply 
or the presence of various toxic chemicals appears to be responsive 
switching. In this paper, we will present a generic model that makes 
it possible to quantify the effect of biochemical noise on the growth 
rate of a clonal population of cells that use this mechanism to 
respond quickly to changes in the environment. 

Our model integrates a description of how the internal dynamics of the
composition of a cell affects the growth rate of that cell with a
description of how the growth rates of the individual cells
collectively determine the growth rate of the population. This allows
us to address a number of fundamental questions: a) How does the
growth rate of the population depend upon the growth rate of a single
cell as a function of its protein expression levels? b) How does the
population's growth rate depend upon the variance and the correlation
time of these fluctuations?  Our model predicts that an important
parameter that controls the effect of biochemical noise is the
correlation time of the fluctuations: only when the correlation time
is long compared to the cell cycle time, does biochemical noise affect
the growth rate of the population.  Interestingly, recent experiments
on {\em E. coli} \cite{Rosenfeld05} and human cells
\cite{Sigal:2006uq} have revealed that the correlation times of
protein concentration fluctuations can be on the order of the cell
cycle time, or even longer.
 Our analysis 
thus predicts that biochemical noise can significantly effect the 
growth rate of a population of cells. Moreover, our model predicts 
that fluctuations can both enhance and reduce the population's growth 
rate. When the average expression level of a protein is close to its 
optimum, fluctuations in its concentration will reduce the 
population's growth rate.  However, when it is sufficiently far from 
its optimal level, 
fluctuations can actually enhance the growth rate of the 
population. This effect arises at the population level and is a 
consequence of the fact that cells that happen to growth faster due to 
noise, become overrepresented in the population. 

Our analysis highlights the difference between ensemble averages and
time averages~\cite{Lu07}.  The ensemble or population average of a
quantity such as protein noise is defined as the average of that
quantity over the cells in the population at a given moment in time;
when a large population exhibits stationary growth, this average does
not change with time.  The time average of a quantity is defined as
the average of that quantity in a single cell and its descendants over
time.  The time average is a property of the intracellular biochemical
network: its value only depends upon the dynamics of the
protein concentrations. In contrast, in experiments often
the ensemble average is measured \cite{Elowitz02,Ozbudak02,Raser05,Kaern05}. Our analysis elucidates that the
ensemble average of a quantity not only depends upon the dynamical properties
of the network, but also on whether fluctuations of this quantity
couple to the growth rate of the cells.

The model also allows us to perform a cost-benefit analysis of 
regulatory control. Recently, Dekel and Alon performed a series of 
experiments that strongly suggest that protein expression is the 
result of a cost-benefit optimization problem \cite{Dekel05b}. They 
showed that the expression level of the \emph{lac} operon is determined by 
the trade off between the cost of synthesizing the metabolic enzyme 
LacZ and the benefit this enzyme confers in enabling the consumption 
of the sugar lactose. In particular, they developed a cost-benefit 
analysis that allowed them to successfully predict the optimal {\em 
  average} expression level of the operon as a function of the lactose 
concentration. However, this analysis does not answer the question how 
the growth rate depends upon the fluctuations in the expression level 
of the metabolic enzyme, nor does it answer the question what 
determines the optimal average expression level of the gene regulatory 
protein that regulates the expression level of the metabolic 
enzyme.

While the cost function of synthesizing a gene regulatory protein is 
probably similar to that of producing a metabolic enzyme, their benefit 
functions are fundamentally different. 
 The benefit of 
producing a metabolic enzyme is that it allows the uptake of the sugar 
by the metabolic network. In contrast, the benefit of synthesizing a 
regulatory protein is indirect and is derived from that of the 
metabolic enzyme; synthesizing a regulatory protein can be beneficial 
because it allows the cell to adjust the expression level of the 
metabolic enzyme to its optimum in response to a changing sugar 
concentration. However, a given optimal expression level of the 
metabolic enzyme as a function of the sugar concentration, does not 
uniquely determine the optimal expression level of the regulatory 
protein. A given optimal response function of the enzyme expression 
level as a function of the sugar concentration, can be obtained by 
different combinations of parameters such as the binding affinity of 
the inducer to the regulatory protein, the binding strength of the 
regulatory protein to the DNA, the degree to which these molecules 
bind cooperatively with each other, as well as the total concentration 
of the regulatory protein (see Figure \ref{fig:Regulation}). What 
determines the optimal combination of these parameters that all can 
yield the same response curve of the enzyme expression level as a 
function of sugar concentration? 

We conjecture that the benefit function of the regulatory protein is 
determined by the fluctuations in the expression level of its target, 
the metabolic enzyme, although other factors such as the response time 
could play a role as well. As we will show, when the average 
expression level of the metabolic enzyme is close to its optimum, 
fluctuations will tend to reduce the population's growth 
rate. Different gene regulatory networks can yield the same average 
response function, but can have markedly different noise 
properties. In particular, our analysis predicts that the inducer, 
e.g. sugar, should bind the gene regulatory protein 
strongly. Moreover, it predicts that higher expression levels of the 
regulatory protein lower the noise in the expression level of the 
metabolic enzyme. We therefore predict that the optimal expression 
level of a regulatory protein is determined by the interplay between 
the cost of making the regulatory protein and the benefit of reducing 
the fluctuations in the target gene. Recently, a similar idea has 
independently been proposed by Kalisky, Dekel and Alon 
\cite{Kalisky:2007fj}. Using as inputs the cost and benefit functions 
as measured by Dekel and Alon \cite{Dekel05b}, our model predicts that 
the optimal expression level of the \emph{lac} repressor should be on the 
order of 10-50 copies, which is remarkably close to the level 
found {\em in vivo} \cite{Gilbert66}.\\ 

\noindent {\bf \Large Results}\\
\noindent {\bf Growth rate} 

In order to describe the effects of biochemical noise on the growth 
rate of a population of cells, we have to develop a model that 
describes how a) the internal dynamics of a cell affects the growth 
rate of that cell and b) how the latter affects the growth rate of the 
population of cells. We now first discuss the latter. 

\noindent{\bf \sf The growth rates of single cells and the 
  growth rate of the population} 
In order to quantify the growth rate of a cell, we have to define a parameter 
that monitors the progress along the cell cycle. This parameter, $Z$, 
could be the amount of replicated DNA, the length of the cell, or a 
combination of these parameters. It has a value $Z=Z_{\rm i}$ at the 
beginning of the cell cycle and a value $Z=Z_{\rm f}$ at the end of 
the cell cycle. The value of the `cell cycle coordinate' $Z$ thus 
exhibits an oscillatory sawtooth pattern as a function of time. Its 
role is analogous to that of a reaction coordinate in chemical 
kinetics, which measures the progress of a chemical reaction and 
serves to define the chemical rate constant. In our case, $Z$ serves 
to quantify the instantaneous growth rate, $\lambda$, of each cell in 
the population: 
\begin{equation} 
\lambda = \frac{dZ}{dt}. 
\label{eq:z} 
\end{equation} 

The growth rate $\lambda$ depends upon the composition of the cell. 
This is determined by the expression level of ribosomal proteins, 
which are needed to make new proteins, and the expression levels of 
metabolic enzymes and other non-ribosomal proteins, which are required 
to produce the building blocks for protein synthesis and cell growth 
\cite{Ingrahambook}. 
 We denote the concentrations of these 
different proteins by $\{X_1,X_2,\dots,X_{n-1},X_n\} \equiv 
{\rm \bf X}$. The growth rate $\lambda$ is thus a function of ${\rm \bf X}$: 
$\lambda \equiv \lambda({\rm \bf X})$. Together with the cell cycle 
coordinate $Z$, ${\rm \bf X}$ specifies the state of each cell in the 
population. 

To determine the growth rate of a population of cells, a key quantity 
is the probability density $P(Z,{\rm \bf X},t)$ to find a cell with a certain 
state $Z,{\rm \bf X}$, inside the population. The 
evolution of this probability density can be expressed in operatorial form as 
\beq  
\frac{\partial  P(Z,{\rm \bf X},t)}{\partial t} = \left[ 
 -\frac{\partial}{\partial Z}  \lambda({\rm \bf X})+ 
\widehat{H}_{\rm X}  - g(t)\right]P(Z,{\rm \bf X},t). 
\label{eq:timeev} 
\eeq The first term on the right-hand side describes the 
evolution of $P(Z,{\rm \bf X},t)$ due to the deterministic evolution 
of $Z$ (see Equation \ref{eq:z}); it corresponds to a Fokker-Planck operator~\cite{Risken96} in the 
limit of zero noise. The operator $\widehat{H}_{\rm X}$ is the 
Fokker-Planck operator encoding the evolution of $P(Z,{\rm \bf X},t)$ 
resulting from the noisy dynamics of the composition ${\rm \bf 
  X}$.  The last term describes the effect of cell 
division on the probability density $P(Z,{\rm \bf X},t)$. Indeed, the 
cell division at $Z_{\rm f}$ amounts to a ``dilution'' of the 
probability of finding cells with intermediate $Z$ values.  The 
steady-state probability distribution function, $P_{\rm s} (Z,{\rm \bf 
  X},t)$, satisfies the equation \beq 0 = \left[ 
  -\frac{\partial}{\partial Z} \lambda({\rm \bf X})+ \widehat{H}_{\rm 
    X} - g\right]P_{\rm s}(Z,{\rm \bf X},t), 
\label{eq:ss} 
\eeq with the boundary condition \beq 2 P_{\rm s}(Z_{\rm f},{\rm \bf 
  X},t_{\rm f}) = P_{\rm s}(Z_{\rm i},{\rm \bf X},t_{\rm i}).  \eeq 
This condition formalizes the observation that upon cell division a 
cell at the end of the cell cycle gives birth to two newborns. 
Importantly, $g$ {\em is} the growth rate of the population of cells 
in steady state. In this ``stationary state'', the number of cells in 
the population grows exponentially, but the 
fraction of cells $P(Z,{\rm X})$ with internal states $Z,{\rm X}$ has 
converged to a time-invariant quantity. 
At each moment in time, there is a constant fraction of cells ready to 
undergo cell division; the number of cells undergoing cell 
division thus grows exponentially with time, but remains proportional 
to the population size, with the proportionality factor given by the 
growth rate $g$. 

\noindent{\bf \sf The growth rates of single cells and protein 
  concentration fluctuations} 
The above model is a generic model of the cell cycle. To make further progress, 
we have to specify the dynamics of ${\rm \bf X}$. The copy number of a 
protein will increase as the cell grows, and will (on average) be 
divided in half when the cell divides. The copy number will thus 
exhibit an oscillatory temporal profile. The volume of the cell will 
show similar oscillatory dynamics. These oscillations will tend to 
cancel each other in their ratio, the {\em concentration} of the 
protein. We make the simplifying assumption that the concentration of 
each species fluctuates around a {\em constant} steady-state level 
during the cell cycle, and that the amplitude of these fluctuations is 
small. It allows us to linearize the interactions between the 
different species at steady state, and to use the linear-noise 
approximation \cite{Elf03}; a comparison with a description based on 
the chemical master equation has shown that this approximation is 
surprisingly accurate, even when the copy numbers are as low as ten \cite{Tanase06a,Ziv07}.  It yields the 
following set of chemical Langevin equations: \beq \dot x_i = 
-\sum_{j=0}^n f_{ij} x_j + \eta_i, \quad \forall i.  \label{eq:langevin}\eeq Here, $x_i = 
X_i - X_{{\rm s},i}$ is the deviation of the concentration $X_i$ of 
species $i$ away from its steady-state value $ X_{{\rm s},i}$, and 
$f_{ij}$ corresponds to the coupling between species $i$ and $j$. The 
term $\xi_i$ describes the noise in $x_i$ that arises from the 
stochastic character of the chemical reactions. We model it as 
Gaussian white noise, with zero mean and variance determined by the 
concentrations of the species at steady state. In Equation 
(\ref{eq:timeev}), the relevant probability density now becomes 
$P(Z,{\rm \bf x},t)$ and the operator that describes the evolution of 
$P(Z,{\rm \bf x},t)$ due to the Langevin dynamics of ${\bf x}$, 
becomes $\widehat{H}_{\bf x}$ (see {\em Methods}). 

If the composition of the cells would not fluctuate in time, then the 
evolution of the cell cycle parameter $Z$ would be deterministic. The 
growth rate $\lambda({\rm \bf X})$ of each cell would then be constant 
in time, $\lambda({\rm \bf X}) = \lambda_0$, and proportional to the growth 
rate of the population, $\lambda_0 \sim g$. In the presence of 
biochemical noise, the growth rate not only depends upon the average 
protein levels, ${\rm \bf X}$, but also upon the fluctuations 
around the average, ${\rm \bf x}$, which lead to variations in the 
growth rate.  It is conceivable that the growth machinery responds 
slowly to fluctuations in the composition in the cell; the growth rate 
would then ``average'' over fluctuations in the composition over some 
characteristic time scale $\tau$: $\lambda =\lambda({\rm {\bf 
  X}_s}, \overline{{\rm \bf x}}^\tau)$, where the bar with the 
superscript $\tau$ indicates that the fluctuations in ${\rm \bf x}$ 
are averaged over a time $\tau$. However, experiments have revealed 
that protein concentrations fluctuate fairly slowly: for {\em 
  E. coli}, the correlation time is on the order of 45 min, which is 
on the order of the cell cycle time \cite{Rosenfeld05}. 
 We argue that since the 
protein concentrations relax slowly, it is reasonable to assume that 
the instantaneous growth rate depends upon the instantaneous 
composition of the cell. We therefore conjecture that the growth rate 
is given by $\lambda =\lambda({\rm {\bf X}_s},{\rm \bf x})$. 

To obtain the growth rate $\lambda({\rm {\bf X}_s},{\rm \bf 
  x})$, we expand it around the steady state ${\rm {\bf X}_s}$ to second 
order in ${\rm \bf x}$ \beq \lambda({\rm {\bf X}_s}, {\rm 
  \bf x})=\lambda_0({\rm {\bf X}_s}) +\sum_{i} a_i x_i + 
\sum_{ij}b_{ij}x_i x_j. 
\label{eq:lambda_x} 
\eeq 
 The 
equation for the stead-state probability density $P_{\rm s} (Z,{\rm 
  \bf x},t)$, Equation (\ref{eq:ss}), can now be solved by making a 
multidimensional Gaussian Ansatz for $P_{\rm s}(Z,{\rm \bf x},t)$ \beq 
P_{\rm s}(Z,{\bf x})\sim 2^{\frac{Z-Z_{\rm i}}{Z_{\rm i}-Z_{\rm 
      f}}}\rme^{-\frac{1}{2}\sum_{ij}\alpha_{ij}(x_i-x^0_i)(x_j-x^0_j)}. 
\label{eq:ans} 
\eeq From now on we shall rescale the time and the $Z$ coordinate such 
that $Z_{\rm f}-Z_{\rm i}= \log(2)$.  In order to understand why such 
a transformation is useful, it should be noted that in the absence of 
protein concentration fluctuations, each cell in the population needs 
a constant time between birth and division $T_{\rm cycle}=(Z_{\rm 
  f}-Z_{\rm i})/\lambda_0$. At the population level, $T_{\rm cycle}$ 
is also the time it takes for the population to double in size, such 
that the growth rate of the population is $g=\log(2)/T_{\rm cycle}$. 
Clearly, in the zero fluctuation limit, the growth rate of the 
population of cells equals the growth rate of each single in the 
population: $g=\lambda_0$. In the presence of protein concentration 
fluctuations, however, the cell cycle times of the individual cells 
will fluctuate, such that even a population of cells that are 
initially perfectly synchronized will eventually converge towards a 
steady-state distribution as given by Equation (\ref{eq:ans}).\\[0.3cm] 

\noindent{\bf Time averages do not always equal ensemble averages}\\ 
Our model shows that the ``time average'' of a quantity such as 
the average protein expression level or the noise in gene expression, 
is, in general, not equal to its ``ensemble average''~\cite{Lu07}.
 The time 
average of a quantity $X$, $\overline{X}$, is defined as the temporal 
average of $X$ along one ``line of descent'': 
\begin{equation} 
\overline{X} = \frac{1}{T} \int_0^T X(t).
\end{equation} 
Here, $X(t)$ can be obtained by monitoring $X$ as a function of time 
in a given cell, whereby upon cell division one follows a randomly 
chosen descendant. The integration time $T$ should be much  
longer than the correlation time of the fluctuations in $X$.  
To obtain better statistics, one 
could average over different trajectories  $X(t)$  in a population,  
but each such path  
has to have a different ancestor (the first cell on the path). 
 The ensemble average of the quantity $X$, $\langle 
X \rangle$, is defined as the average of $X$ across the population of 
cells: 
\begin{equation} 
\langle X \rangle = \frac{1}{N(t)}\sum_{\alpha=0}^{N(t)} X_\alpha(t), 
\end{equation} 
where $N(t)$ is the number of cells in the population at time $t$ and 
$X_\alpha(t)$ is the magnitude of $X$ in cell $\alpha$ at time $t$; 
when the growing population is in the stationary state and $P(Z,{\rm 
  \bf X},t)$ is time invariant, this ensemble average does not change 
with time. To illustrate the difference between the two kinds of 
averages, let's consider the fluctuations in the composition ${\rm \bf 
  X}$. To the extent that protein concentration fluctuations are 
described by the chemical Langevin equation (Equation 
\ref{eq:langevin}), the distribution of the concentrations ${\rm \bf 
  X}$ as obtained by following the time traces of $X_i$ in a given 
cell and its descendants, is given by a Gaussian that is centered at 
$\overline{{\rm \bf X}}={\rm \bf X}_{\rm s}$. In contrast, the 
distribution of ${\rm \bf X}$ over different cells in a population at 
a given moment in time is also a Gaussian, but now the Gaussian is 
centered at $\langle {\rm \bf X} \rangle = {\rm \bf X}_{\rm s} + {\rm 
  \bf x}^0$, where ${\rm \bf x}^0$ may deviate from zero. Moreover, 
not only the mean, but also the variance of the two distributions 
will, in general, differ, as we will show now. 
\\[0.3cm] 

\noindent{\bf Biochemical noise can both reduce and enhance the 
  population's growth 
  rate}\\ 
 In order to understand the non-trivial 
effects of biochemical noise on the growth rate of a population of 
cells, it is instructive to consider a simple example. Let's consider 
a single metabolic enzyme X, and assume that the temporal dynamics 
of its concentration is given by 
\begin{equation} 
\dot x = -\gamma x + \eta, 
\end{equation} 
where $x$ is the deviation of the enzyme concentration $X$ away from 
its steady-state value, $X_{\rm s}$, $\gamma^{-1}$ is the response 
time, which is typically on the order of the cell cycle time, and 
$\eta$ is a Gaussian white noise term, of zero mean and strength $2 
D$. The time average of the variance of the fluctuations in the 
concentration of X as obtained from the time trace of $X$ of a given 
cell and its descendants, is $\overline{X^2}-\overline{X}^2 = 
\sigma^2_{\rm X} = D/\gamma$. 

We assume that over the concentration range of interest, the growth 
rate of a given cell as a function of the expression level of X can be 
written as 
 \beq \lambda=\lambda_0 (X_{\rm s}) + a x + b x^2,  
\label{eq:lambda_X} 
\eeq 
where $\lambda_0 (X_{\rm s})$ is the growth rate of the cell when 
the enzyme concentration equals $X_{\rm s}$. The growth rate of the 
population of cells is then given by (see {\em Methods}) \beq 
g=\lambda_0 (X_{\rm s}) +\frac{a^2 D}{\gamma^2-4 b D}+b \sigma^2. 
\label{eq:simpleq} 
\eeq Here, $\sigma^2$ is the variance of the fluctuations in $X$ 
within the population of cells at a given time: $\sigma^2 = \langle 
X^2\rangle - \langle X \rangle^2$. This ensemble or population average 
is given by \beq \sigma^2=\frac{2D}{\gamma+\sqrt{\gamma^2-4 b D}}. 
\eeq The ensemble average $\sigma^2$ can be written in terms of the 
time average of the variance, $\sigma^2_{\rm X}$: $\sigma^2 = \frac{2 
  \sigma^2_{\rm X}}{1+\sqrt{1-4b\sigma^2_{\rm X}/\gamma}}$. Clearly, 
if the growth rate is non-linear in $X$, i.e. if $b\neq 0$, the 
ensemble average of the variance in $X$ does not equal its time 
average. Importantly, the time average of the protein noise, 
$\sigma_{\rm X}^2$, is a characteristic of the stochastic properties 
of the underlying biochemical network. However, the protein noise is 
often measured as an ensemble or population average 
\cite{Elowitz02,Ozbudak02,Raser05,Kaern05}. Our results show that  if one is 
interested in the noise properties of the underlying network,  one 
should compute the protein noise by combining sequential noise traces 
of cells through lines of descent \cite{Austin06} when 
the expression of the fluorescent protein used to measure the noise 
affects the growth rate significantly (such that $b$ is much smaller 
than zero). 

Let us now consider the scenario in which the average expression 
level of the enzyme is such that the growth rate is maximal: 
$X_{\rm s}=X_{\rm opt}$ (see Figure \ref{fig:sketch}). In this 
case, $a$ is zero, and $b = \partial^2 \lambda /\partial X^2 < 0$. The 
growth rate of the population is then $g =\lambda_0 (X_{\rm s}) + 
b \sigma^2$. Since $b$ is negative, $g < \lambda_0$. Hence, when the 
composition is close to its optimum, biochemical noise always tends to 
reduce the overall growth of the population. 

If the average expression level $X_{\rm s}$ deviates significantly 
from the optimal expression level $X_{\rm opt}$, the situation is 
qualitatively different (see Figure \ref{fig:sketch}). Sufficiently 
far away from the optimum, the curvature can be ignored ($b=0$), and 
the growth rate is given by $g=\lambda_0+a^2D/\gamma^2 = \lambda_0 + 
a^2 \sigma^2_{\rm X} / \gamma$. In this regime noise always increases 
the growth rate, irrespective of the sign of $a$, and even though at 
the {\em single cell} level the growth rate $\lambda$ is linear in 
$X$.  
The reason is that cells that happen to have a composition that 
is closer to the optimum, will grow faster and therefore divide 
earlier; moreover, the daughter cells will inherit the composition 
from their mother, and will thus also grow faster than the 
steady-state value, and so on. As a consequence, cells with a higher 
growth rate become overrepresented in the population, which can be 
verified by noting that the mean of $x$ in the population of cells is 
now shifted from zero to $x^0 = Da / \gamma^2 = a \sigma^2_{\rm X} / 
\gamma$. This mechanism, whereby the cells that grow faster due to a 
fluctuation in their protein composition generate more off-spring, 
increases the overall growth rate of the population.  The increase in 
the growth rate due to noise, $a^2 \sigma^2_{\rm X} / \gamma$, depends 
upon how strongly the growth rate changes with $X$, which is given by 
the slope $a$, and on the magnitude of the concentration fluctuations 
in each cell, given by $\sigma^2_{\rm X}$.  Importantly, it also 
depends upon the relaxation time of the fluctuations, given by 
$\gamma^{-1}$. If the response time is much faster than the cell cycle 
time, then on the relevant time scale of the cell cycle, the 
concentrations in all the cells will be the same and no benefit from 
the noise can be gained. However, both in prokaryotic 
\cite{Rosenfeld05} and eukaryotic cells \cite{Sigal:2006uq}, 
correlation times of protein concentration fluctuations have been 
measured to be on the order of the cell cycle time or longer, meaning 
that they are potentially important. Please also note that a non-zero 
$x^0$ means that the time average of $X$, which is given by 
$\overline{X}=X_{\rm s}$, is not equal to the ensemble average of $X$, 
which is given by $\langle X \rangle = X_{\rm s} + x^0$. 

Lastly, we note here that it is conceivable that the curvature $b$ of 
the growth rate $\lambda$ is locally positive. In this case, the solution 
to Equation (\ref{eq:simpleq}) is only valid when $\gamma^2 > 4 bD$. At the 
point where this condition is no longer satisfied, an interesting 
bifurcation can arise towards a state where the growth dynamics alone 
imposes a bimodal distribution of protein concentrations: in the 
population, cells with a high expression level then co-exist 
with cells with a low expression level.~\\ 

\noindent{\bf Fluctuating environment}\\ 
The analysis above describes how fluctuations in the composition can 
affect the growth rate of a population of cells in a constant 
environment. We now briefly discuss how fluctuations in the 
environment affect the population's growth rate. As before, we 
consider the scenario in which cells respond to changes in the 
environment via the mechanism of responsive switching: they thus sense 
the changes in the environment and respond appropriately. 

If the environmental signals are described by the vector ${\rm \bf S}$, 
then the time varying environment can, in general, be decomposed as: 
\begin{equation} 
{\rm \bf S} = {\rm \bf S}^{\rm c} + {\rm \bf S}^{\rm u}. 
\end{equation} 
Here, ${\rm \bf S}^{\rm c}$  denote the correlated fluctuations between 
the different cells, while ${\rm \bf S}^{\rm u}$ corresponds to the 
fluctuations in the environmental signals that are uncorrelated from 
one cell to the next within the population.  

The uncorrelated fluctuations in the external signals can be treated 
in the same spirit as the fluctuations in the internal 
signals. Their dynamics could be added to that of ${\bf x}$: 
\begin{eqnarray} 
\dot s^u_i &=& -\mu_i s^u_i + \xi_i, \quad i=\overline{1\dots m},\\ 
\dot x_i &=& -\sum_{j=0}^n f_{ij} x_j + \sum_{j=0}^m g_{ij} s^{\rm u}_j + \eta_i, \,\, i=\overline{1\dots n} , 
\end{eqnarray} 
where $s^{\rm u}_i=S^{\rm u}_i - S^{\rm u}_{\rm s,i}$, with $S^{\rm 
  u}_i$ the part of the fluctuations of the external signal $i$ that 
is uncorrelated between different cells, and $g_{ij}$ indicates how 
the internal dynamics of species $i$ is coupled to the fluctuations in 
the external signal $j$. Since the fluctuations in ${\rm \bf S}^{\rm 
  u}$ couple to the fluctuations in the composition ${\rm \bf X}$, 
they could either reduce or enhance the growth rate of the population, 
depending on whether the composition ${\rm \bf X}$ is close to its 
optimum or not, respectively. 

The effect of the correlated fluctuations in the external signals, 
${\rm \bf S^{\rm c}}$, are much more difficult to treat analytically 
\cite{Thattai04}. However, if these fluctuations occur on a time scale 
that is much longer than the time it takes for the internal dynamics 
${\rm x}$ 
to relax towards a new steady state after an environmental change, 
the overall growth rate can be written as \beq g=\int \rmd {\rm \bf 
  S}^{\rm c} P({\rm \bf S}^{\rm c}) g({\rm \bf S}^{\rm c}). 
\label{eq:avg} 
\eeq  
This expression shows that the cells need to adapt 
to a given distribution of external signals.  

We can make an estimate for the time it takes for the population to 
relax towards a new steady after a change in the environment has 
occurred. If prior to an environmental change, the cell cycle 
coordinate $Z$ has reached steady state, meaning that $P(Z)$ is 
uniform across the population of cells, then $P(Z)$ does not have to 
relax towards a new steady state after the change in the 
environment. The distribution in the composition, $P({\rm \bf X})$, 
however, does have to relax. If the relaxation time of the population 
is dominated by the slow dynamics of a single protein X, the 
relaxation rate is given by $k = \sqrt{(\gamma^2 - 4 bD)}$.  This 
shows that in the absence of fluctuations ($D=0$) the relaxation rate 
is given by the rate of protein decay, $\gamma$, as one would 
expect. It also shows that when the growth rate of a cell is a concave 
function of $X$ ($b < 0)$, fluctuations can actually enhance the 
relaxation rate; the reason is that cells that are closer to the new 
optimum will grow faster. This analysis shows that a conservative 
estimate for the validity of Equation (\ref{eq:avg}) is that the 
environmental fluctuations should occur on time scales longer than the 
protein decay time $\gamma$. 
\\ 

\noindent{\bf The cost of reducing noise: optimal expression levels 
  of gene regulatory proteins}\\ 
In order to understand the design criteria that determine the 
magnitude of the fluctuations in the expression level of a given 
protein for cells that respond via responsive switching, we do not 
only have to understand how these fluctuations affect the growth rate, 
as discussed above, but also the indirect energetic cost of 
controlling these fluctuations. Both the magnitude of the 
concentration fluctuations and the cost of controlling these 
fluctuations are determined by the design of the network that 
regulates the expression level of the protein of interest. We will now 
show, using the \emph{lac} system as an example, that the optimal design of 
the regulatory network is determined by the interplay between these 
two factors.

We use a simple model of the \emph{lac} system in the absence of glucose but 
in the presence of lactose. The inducer lactose (ligand L) binds the 
lac repressor (transcription factor TF); upon binding, the transcription factor 
dissociates from the operator and the enzyme, LacZ in this case, is 
expressed.  We assume that both the binding of ligand to the 
transcription factor and the binding of the latter to the operator are 
fast such that they can be integrated out. The dynamics of 
the regulatory protein and the metabolic enzyme is then specified as: \bea 
\dot x &=& -\gamma x +\xi_{\rm X}, \nonumber\\ 
\dot e &=& -\gamma e + f x + \xi_{\rm E}. 
\label{langevinlac} 
\eea Here, $x$ denotes the deviation away from the total steady-state 
TF concentration, denoted by $X_{\rm s}$, $e$ denotes the 
deviation away from the steady-state concentration of the enzyme, 
$E_{\rm s}$, $\gamma$ is the degradation rate of both proteins, 
and $\xi_{\rm X}$ and $\xi_{\rm E}$ model the (Gaussian white) noise in their 
expression.  The factor $f$ is the differential gain that describes 
the change in the protein production rate (expression rate) $k_{\rm E}(X)$ due to 
a change in the concentration of the transcription factor: 
$f=\partial{k_{\rm E}(X)}/\partial {X}$. In this expression we integrate the 
contributions of TF-ligand binding, TF-operator binding, and the 
dynamics of mRNA.  The fluctuations in $e$ have an intrinsic source, 
modeled by $\xi_e$, and an extrinsic one that arises from the 
fluctuations in $x$. Since the expression level of the enzyme is much 
higher than that of the gene regulatory protein, the dominant source 
of noise in $e$ is the extrinsic one, arising from fluctuations in the 
TF concentration. In what follows, we therefore 
ignore the intrinsic contribution $\xi_{\rm E}$. 

To make further progress, we need to know how the growth rate of each 
cell, $\lambda$, depends upon the expression level of the enzyme and 
that of the transcription factor.  Recently, Dekel and Alon 
\cite{Dekel05b} performed a series of experiments that allowed them to 
measure both the cost and the benefit of producing the metabolic 
enzyme LacZ. By using an artificial inducer, they varied the expression 
level of LacZ in the absence of its substrate lactose, and measured 
the effect on the growth rate.  The inducer induces the production of 
LacZ, but no benefit is gained, since the lactose is absent and the 
inducer is not metabolized. This set of experiments thus allowed them 
to determine the cost of synthesizing the LacZ protein. In a separate 
set of experiments they measured how the growth rate changes with the 
lactose concentration, when the expression level is kept constant (due 
to a saturating amount of the inducer). This set of experiments gave 
them an (indirect) estimate of the benefit function. By assuming that 
the optimal expression level is given by the level that maximizes the 
benefit minus the cost, the measured cost and benefit functions could 
be used to predict the optimal LacZ expression level as a function of 
lactose concentration. 

Following Dekel and Alon \cite{Dekel05b}, we write the change in the 
growth rate of a single cell, $\Delta \lambda = \lambda - \lambda_0$, due to the 
production of the gene regulatory protein and the metabolic enzyme 
relative to the growth rate in the absence of these proteins, 
$\lambda_0$, as:  
\beq 
\label{eq:CB} 
\frac{\Delta \lambda}{\lambda_0}=\delta (E_{\rm s}+e) - \eta 
\frac{(E_{\rm s}+e+X_{\rm s}+x)}{1-\frac{E_{\rm s}+e+X_{\rm 
      s}+x}{M}}.\eeq The first term on the right-hand side encodes the 
gain in the growth rate due to the metabolic activity of the enzyme; 
importantly, $\delta \equiv \delta(L)$ is a function of the lactose 
concentration $L$ (see Equation \ref{eq:eta_delta} below).  The second 
term, with $\eta$ being a constant, quantifies the cost of producing 
the enzyme and the regulatory protein; the factor $M$ is the maximal 
capacity for producing non-essential proteins \cite{Dekel05b}. Note 
that we assume that the costs of producing one enzyme molecule and one 
gene regulatory protein molecule are the same. 

As discussed in the introduction, a given average optimal expression 
curve of $E$ as a function of sugar concentration, $E_{\rm opt}(L)$, 
can be obtained by different expression levels of $X$. A mean-field 
analysis, which ignores the effect of fluctuations in $E$ and $X$, 
would predict that the optimum expression level of $X$ is close to 
zero, since that minimizes the cost of producing the regulatory 
protein. We therefore assume that the steady-state enzyme expression 
level, $E_{\rm s}$, is given by that level $E_{\rm opt}^0$ that 
maximizes $\Delta \lambda$ with respect to $E$ {\em at} $X = 0$. The 
steady-state enzyme expression level is thus given by 
 \beq E_{\rm 
  s}=E_{\rm opt}^0 = 
M\left(1-\sqrt{\frac{\eta}{\delta}}\right).\label{eq:Eopt}  
\eeq This 
expression is, in fact, the principal result of the cost-benefit 
analysis of the optimal enzyme expression level of Dekel and Alon 
\cite{Dekel05b}. The expression, with $\delta$ being a function of the 
lactose concentration (see Equation \ref{eq:eta_delta} below), gives a 
remarkably good prediction for the enzyme expression level as a 
function of the lactose concentration \cite{Dekel05b}. The prediction 
is shown in Figure \ref{fig:Regulation}C. We now address the 
question what is the optimal regulatory network---the optimal 
TF concentration $X_{\rm s}$, the optimal TF-L and 
TF-operator binding strengths---under the assumption that the 
steady-state enzyme expression level as a function of lactose 
concentration is fixed 
and given by Equation (\ref{eq:Eopt}): $E_{\rm 
  s} (L) = E_{\rm opt}^0 (L)$. 

To obtain the growth rate at $E=E_{\rm s} + e$ and 
$X=X_{\rm s} + x$ (with finite $X_{\rm s}$), we expand the growth rate 
around $E_{\rm opt}^0$ and $X=0$, which yields the following 
expression for the relative growth rate (see {\em Methods}): 
 \beq \frac{ 
  g-\lambda_0}{\lambda_0}=M(\sqrt{\delta}-\sqrt{\eta})^2 - 
\delta X_{\rm s} -\frac{\delta}{2 M} 
\sqrt{\frac{\delta}{\eta}}\frac{f^2}{\gamma^2}\sigma_{\rm X}^2. 
\label{eq:final} 
\eeq On the left-hand side of the above equation, $g$ is the growth 
rate of the {\em population} of cells. The first two terms on the 
right-hand side give the deterministic, mean-field prediction that 
ignores the effect of fluctuations in $x$ and $e$: in the absence of 
fluctuations, the growth rate of the {\em population} of cells, $g$, 
equals the growth rate of each single cell, $\lambda_{\rm D}$, which is 
given by $\lambda_{\rm D}=\lambda_0+\lambda_0 \left [(\sqrt{\delta} - 
  \sqrt{\eta})^2M - \delta X_{\rm s})\right]$ (see {\em Methods}). The 
last term of Equation (\ref{eq:final}) describes the effect of fluctuations 
on the growth rate. The second term on the right hand side shows that at the 
mean-field level, there is indeed a pressure to minimize the 
production of the regulatory protein X; this is associated with 
minimizing the cost of producing the regulatory protein. The third 
term on the right hand side shows, however, that there is also a pressure to 
minimize the fluctuations in X, given by $\sigma_{\rm X}^2$. Its 
origin is that fluctuations in the gene regulatory protein $X$ lead to 
fluctuations in $E$, and since the mean expression level of $E$ is 
assumed to be at its optimum, these fluctuations tend to lower the 
growth rate. Importantly, the magnitude of the fluctuations in $X$ and 
hence $E$ decreases as the average expression level of $X$ 
increases. Clearly, while the cost of producing $X$ tends to lower the 
optimal expression level of $X$, the benefit of reducing the 
fluctuations in $E$ tends to increase the optimal expression level of 
$X$. The optimal expression level of $X$ is determined by the balance 
between these two opposing factors. A similar conclusion was 
recently independently reached by Kalisky {\em et al.} \cite{Kalisky:2007fj}. 

To demonstrate this explicitly, we will study in more detail the last 
two terms in Equation (\ref{eq:final}), which describe the contribution of the 
transcription factor to the growth rate: \beq 
-\delta\left(\frac{1}{M} 
  \sqrt{\frac{\delta}{\eta}}\frac{f^2}{\gamma^2}\sigma_{\rm X}^2+X_{\rm 
    s}\right). 
\label{eq:partial} 
\eeq 
In our model, the 
steady-state enzyme concentration is given by $E_{\rm s} = E_{\rm opt} = 
k_{\rm E}(X_{\rm s},L)/\gamma$, which means that the gain is given by 
\beq 
 \frac{f}{\gamma} = \frac{\partial{E_{\rm s}}}{\partial {X_{\rm s}}} \simeq \frac{E_{\rm 
   s}}{X_{\rm s}}.  
\label{eq:fog} 
\eeq  
To make further progress, we have to assume a model for the 
fluctuations in $X$.  If we assume that these fluctuations are 
Poissonian, then $\sigma^2_{\rm X} \simeq N_{\rm X}/V^2$ \cite{Ozbudak02}, 
where $V$ is the volume and $N_{\rm X}$ is the copy number of 
$X$. Recent results show that while the fluctuations can be stronger 
than Poissonian due to, for example, bursts in gene expression, the 
linear scaling of $\sigma^2_{\rm X}$ remains correct for many proteins in 
prokaryotes \cite{Bar-Even06}.  Finally, if we assume that $E_{\rm 
  s} \propto M$, the expression in Equation (\ref{eq:partial}) is proportional 
to  
\beq  
-\left( \frac{N_{\rm E}}{N_{\rm X}}+ {N_{\rm X}}\right). 
\label{final2} 
\eeq  
This expression shows a maximum as a function of $N_{\rm X}$. The 
position of this optimum---the copy number of X that maximizes the 
growth rate---is related to the copy number of E by 
\beq  
\label{eq:N_X_N_E}
N_{\rm X}\propto\sqrt{ N_{\rm E}}.  \eeq We therefore predict that the
optimal TF copy number is linear in the square root of the copy number
of the enzyme it regulates. This prediction could perhaps be tested by
performing a statistical analysis of the expression levels of
transcription factors and the expression levels of the target genes
these transcription factors regulate. Such a statistical analysis
could be performed in the spirit of that of Ref. \cite{Bar-Even06}, in
which the authors studied the variation in the expression levels of 43
{\em Saccharomyces cerevisiae} proteins, in cells grown under 11
experimental conditions. Our analysis would predict that if one would
measure the expression levels of transcription factors and their
target genes in such an experiment, the two would be correlated
according to Equation (\ref{eq:N_X_N_E}).

Dekel and Alon \cite{Dekel05b} measured the 
quantities $\delta$ and $\eta$ used above (Equation \ref{eq:CB}) for the \emph{lac} system: 
 \beq  
\eta=0.02 E_{\rm WT}^{-1}, \quad \delta=0.17 
E_{\rm WT}^{-1} \frac{L}{0.4{\rm mM}+L}, 
\label{eq:eta_delta} 
\eeq  
(where $L$ is measured in 
${\rm mM}$ units). Here $E_{\rm WT}$ is the fully induced wild-type concentration of the enzyme, and 
we use $M=1.8E_{\rm WT}$. 
 As explained in the section \emph{Fluctuating 
  Environment} the growth rate in a slowly fluctuating environment 
can be obtained as an average over the different levels of the lactose 
in the environment. As we do not know the wild type distribution of 
sugar the bacterium experiences, we use either a uniform distribution 
over all possible lactose levels in the interval 0-6mM or a 
non-uniform bimodal one that peaks at small and high lactose concentrations.

Figure \ref{fig:unifvsnonunif} shows the optimal repressor expression level, 
for the two different lactose distributions in the environment.  It is 
seen that the growth rate as a function of the copy number of the 
regulatory protein exhibits a broad optimum at around 10-50 
molecules. Interestingly, this is in the biological range \cite{Gilbert66}. Even though 
our model of gene expression is rather simplified (we use, e.g., a constant 
amplification factor $f$), it appears that the prediction of our model 
is remarkably accurate. Interestingly, Kalisky {\em et al.} 
arrived at a similar prediction, even though their model differs in a 
number of ways from ours, as discussed in more detail in the {\em 
  Discussion} section \cite{Kalisky:2007fj}. 

Equation (\ref{eq:final}) shows that the effect of the noise in $X$, 
$\sigma_{\rm X}^2$, on the fluctuations in ${\rm E}$, and hence on the 
growth rate, is determined not only by the decay rate $\gamma$, which 
controls the extent to which fluctuations in $X$ and $E$ lead to 
significant differences between cells in their composition on the time scale of the 
cell cycle, but also by the gain $f$, which determines the extent to 
which the fluctuations in $X$ are amplified.  As we will show now, the 
optimal TF-ligand binding curve and TF-operator binding curve is 
determined by the requirement that the gain $f$ should be minimized as 
much as possible. Let's imagine that the binding of the ligand $L$ to 
the repressor $X$ is given by \beq X_{\rm free}=\frac{X K_{\rm D}}{K_{\rm D}+L}. 
\label{eq:XfL} 
\eeq 
 Here, $X$ is the total TF concentration, $X_{\rm free}$ is 
the concentration of $X$ that is not bound to the inducer, and $K_{\rm D}$ 
is the dissociation constant for ligand-TF binding. The unbound 
transcription factor 
represses the expression of $E$ via the repression function $R \equiv 
R(X_{\rm free}(X,L))$, given by 
 \beq 
E_{\rm opt}(L)=\frac{k_{\rm E}(X)}{\gamma}=\frac{R[X_{\rm free}(X,L)]}{\gamma}.  
\eeq

We 
show these relations in Figure \ref{fig:Regulation}. It is important to note 
that the repression function $R(X_{\rm free}(X,L))$ is not necessary a simple 
Hill function; in the \emph{lac} system this curve is known to be implemented 
with a complicated cooperative interaction and binding to multiple 
operator sites on DNA. 
 Using Equation (\ref{eq:fog}), $\partial{E_{\rm s}}/\partial {X_{\rm s}} =\frac{\partial{E_{\rm 
       s}}}{\partial L}{\frac{\partial L}{\partial 
     X_{\rm free}}}\frac{\partial{X_{\rm free}}}{\partial {X_{\rm s}}}$ and 
 Equation (\ref{eq:XfL}), we arrive at 
\beq 
\frac{f}{\gamma}=-\frac{\partial{E_{\rm opt}}}{\partial L}\frac{ (K_{\rm D}+L)}{X}. 
\label{eq:ldep} 
\eeq To minimize the gain $f$, and hence the effect of noise in $X$ on 
the growth rate, $K_{\rm D}$ should be as small as possible, which 
corresponds to strong TF-L binding. Since the function $E_{\rm 
  opt}(L)$ is assumed to be fixed, strong TF-L binding also implies 
strong TF-operator binding. Hence, as long as TF-ligand binding and 
TF-operator binding can be integrated out, the best strategy would be 
strong TF-L and TF-operator binding. This is illustrated in Figure 
\ref{fig:contourplot}, which shows for the \emph{lac} system the 
contour plot of the 
optimal growth rate in the plane ($X$, $K_{\rm D}$). The conclusion 
that TF-L and TF-operator binding should be strong is supported by the 
experimental observation that the dissociation constant for the 
binding of lac 
repressor to its primary operator site is in the nM range, while the 
binding of the inducer allolactose to the repressor is on the order of 
0.1 $\mu{\rm M}$  \cite{Yagil71}.\\

\noindent{\bf \Large Discussion}\\ 
The response machinery allows a living cell to adjust its composition 
to a changing environment. If the response machinery is fast and 
operates well, then in each environment the cell's composition is 
optimized such that the growth rate is maximized. Our analysis 
suggests that under these conditions, there is an evolutionary 
pressure to minimize the fluctuations in the composition. However, the 
response machinery cannot always optimally adjust the cell's 
composition. When there is a drastic change in the environment, for 
instance, the cell probably has to change its genotype so as to change 
its response machinery.  Our analysis suggests that along such an 
``evolutionary trajectory'' from a sub-optimal configuration of the 
response machinery to a new optimal one, fluctuations in the 
composition could be beneficial, because cells that happen to have a 
composition that is closer to the new optimum will grow more rapidly 
and thereby increase the overall growth rate of the population. 
Based on this 
observation we predict that the periods of fast evolution (for example 
when a population colonizes an entirely new environment) are 
correlated with a positive influence of fluctuations and thus an 
increased variability in the population.  This idea is supported by 
the observation that the regulatory networks that control the response 
to environmental changes are in general noisier than the conserved 
cell machinery \cite{Newman06, Bar-Even06}. 

It has been recognized before in a different context that phenotypic 
variance can be detrimental under stabilizing selection for the 
optimal genotype and advantageous far from this optimal genotype 
\cite{Gavrilets94,Zhang05}. Moreover, it has been suggested that 
phenotypic variance could be maintained if there is an ``engineering'' 
cost of minimizing fluctuations \cite{Zhang05}. Our model, however, 
makes it possible to make a quantitative prediction on the effect of 
protein concentration fluctuations on the growth rate of a clonal 
population of cells. In 
particular, the model predicts that the effect of fluctuations in the 
concentration of a given protein X depends upon the following 
quantities (see Equation \ref{eq:simpleq}): a) the growth rate of a 
single cell as a function of the expression level, $\lambda(X)$ 
\cite{Dekel05b}; b) the strength of the fluctuations in $X$, 
$\sigma_{\rm X}^2$; c) the correlation time of the fluctuations in $X$, 
given by $\gamma$. All these quantities can be measured 
experimentally, which would allow for a quantitative test of our model. In this 
respect, it would be of particular interest to investigate one of the 
key ingredients of our model, which is how the growth rate of a single 
cell, $\lambda$, depends on the composition ${\bf \rm X}$. We have 
assumed that the growth rate depends upon the instantaneous 
composition, but it is conceivable that the growth rate responds to 
changes in the composition with a time lag; alternatively, it could 
depend upon the composition as averaged over some time scale $\tau$: 
$\lambda = \lambda(\overline{\rm \bf X}^\tau)$. 

Recently, Kalisky, Dekel and Alon \cite{Kalisky:2007fj} reported an 
analysis of the optimal design of the gene regulatory network that 
controls the expression of the \emph{lac} operon, which complements 
ours. While we assume that the correlated fluctuations in the 
environment are slow, they also consider correlated fluctuations in 
the environment that are relatively fast to the response time; on the 
other hand, their analysis does not address the question of the 
optimal dissociation constants for inducer-TF and TF-operator 
binding. Our analyses also differ in the description of the extrinsic 
contribution to the noise in the expression of the \emph{lac} operon, 
and in the estimate of the burst size of lac expression. More 
importantly, Kalisky {\em et al.} used a simpler model to describe the 
effect of biochemical noise on the growth rate of a population of 
cells. Our model integrates a description of the effect of noise on 
the growth rate of a single cell with a description of how the growth 
rates of the single cells collectively determine the growth rate of 
the population. In contrast, their model assumes that the growth rate 
of the population is given by the average of the growth rates of the 
individual cells. This approximation does not allow the 
model of Kalisky {\em et al.} to predict that the noise can 
also enhance the growth rate of the population. This is indeed an 
effect that arises at the population level; it is a consequence of the 
fact that cells that happen to grow faster will take over the 
population. Moreover, our work illustrates the importance of the 
correlation time of the protein concentration fluctuations. However, 
the present work agrees with that of Kalisky, Dekel and Alon 
\cite{Kalisky:2007fj} in that we both find that the optimal 
concentration of a gene regulatory protein is determined by the 
interplay between the cost of synthesizing the regulatory protein and 
the benefit of reducing the fluctuations in the expression of its 
target gene. Even quantitatively, the predictions of our models for 
the optimal \emph{lac} repressor concentration are fairly similar, 
although the model presented here would predict a slightly lower 
optimum concentration and a slightly smaller change in growth rate for 
deviations away from this optimum; this could be due to our 
conservative estimate of the burst size.

Our model predicts that if the expression level of the gene regulatory
protein is varied by a factor 2 from its optimal value, the change in
the growth rate would be on the order of $10^{-4}$. This change is
sufficient to provide a selection pressure that is large enough in a
typical bacterial population with an effective size larger than $10^6$
cells; indeed, as discussed in \cite{Wagner05}, relative growth rate
changes as low as $10^{-6}$ are sufficient to balance the genetic
drift in such a population. A change in the growth rate of $10^{-4}$
is thus large enough to provide a selection mechanism in a typical
bacterial population for driving the transcription factor expression
level to within a factor 2 from the predicted optimal level.

Another fundamental question we can address with our model is the 
relative efficiency, from the fluctuations point of view, of different 
modes of regulation (see {\em Methods}). For example, the cost-benefit 
function of Dekel and Alon implies that 
the cost grows with a linear combination of the total enzyme and 
transcription factor concentration, with positive coefficients 
\cite{Dekel05b}. As a consequence, regulatory networks with 
anticorrelated fluctuations of the enzyme and TF concentrations, which 
correspond to repressor based regulatory networks, will provide an 
advantage over those with correlated fluctuations, as for activator 
based regulatory networks. This result is consistent with the 
observation that simple organisms have more repressors than 
activators. Unlike alternative explanations for this observation based 
on the requirement for genotypic robustness with respect to mutational 
fluctuations \cite{Savageau74,Savageau77}, our explanation does not 
require that the rate of environmental fluctuations is comparable to 
the slow relevant mutation rates. 

In this paper, we have focused on the expression of a single 
protein. Yet, it is clear that the model presented in {\em Growth 
  rate} could be used to study more complicated networks as well. In 
these networks, the propagation of noise 
\cite{Detwiler00,Paulsson04,Pedraza05,Shibata05} and hence the effect 
of noise on the growth rate, can be intricate, especially when there 
are (anti-) correlations between different sources of noise 
\cite{Tanase06a,Levine07}.  The model could also be 
used in conjunction with partial-differential equation solvers to 
study non-linear networks, for which biochemical noise is expected to 
become even more important. 

How could our predictions be tested experimentally? Ideally one would like 
to perform an experiment in which the {\em average} expression level 
of the metabolic enzyme is fixed, while the noise in the expression 
level is varied. Several strategies could be envisioned. First of 
all, one could vary the noise level by playing with the transcription 
and translation efficiencies \cite{Ozbudak02,Paulsson04}. To make more 
direct contact with the predictions presented here, however, it would 
perhaps be more interesting to vary the expression level of the 
regulatory protein, while simultaneously varying the TF-operator 
binding strength such that the average expression level of the 
metabolic enzyme remains constant. Alternatively, one could vary the 
expression level of the regulatory protein, while simultaneously 
changing the concentration of an artificial inducer such that the 
enzyme concentration remains constant.  For example, it is possible to 
increase the binding affinity of the \emph{lac} repressor to the 
operator, and therefore the repression strength by a factor as high as 
10, by either mutating the repressor LacI \cite{Kolkhof92} or 
the operator sites  \cite{Sadler83}.  Our analysis predicts that the 
growth rate as a function of the expression level of the regulatory 
protein exhibits a broad maximum as shown in 
Figure \ref{fig:unifvsnonunif}.\\

\noindent{\bf \Large Methods}\\ 
{\bf The stationary distribution $P_{\rm s}(Z,{\bf x})$}\\ 
In this section we derive the solution (Equation \ref{eq:ans}) for the 
stationary probability distribution $P_{\rm s}(Z,\bol{x})$. The 
equation satisfied by $P(Z,\bol{x},t)$ for the case of linear Langevin 
dynamics is: 
\begin{widetext} 
\bea 
\frac{\partial P}{\partial t}&=& -\frac{\partial (\lambda P)}{\partial Z} - g(t) P+\sum_{i}\left[\sum_j\left( D_{ij} \frac{\partial^2 P}{ \partial x_i \partial x_j }+ 
\frac{\partial ( f_{ij} x_j P)}{\partial x_i}\right)\right]. 
\label{eq:fp} 
\eea The three terms on the right hand side of Equation (\ref{eq:fp}) 
describe, in order, the drift along the cell-cycle coordinate $Z$, the 
normalization of $P$ due to the continuous birth 
of new cells in the population, and the Fokker-Plank operator 
describing the internal dynamics of the composition of the individual 
cells \cite{Gillespie00,Risken96}. The diffusion strength $D_{ij}$ is given by 
$\langle\eta_{i}\eta_j\rangle=2 D_{ij}$, where $\langle \eta_i 
\eta_j\rangle$ are the cross-correlations in the Gaussian white noise 
of $X_i$ and $X_j$ \cite{Tanase06a}.  The stationary solution 
satisfies the equation: \bea 0&=& -\frac{\partial (\lambda P_{\rm 
    s})}{\partial Z} - g P_{\rm s}+\sum_{i}\left[\sum_j\left( D_{ij} 
    \frac{\partial^2 P_{\rm s}}{ \partial x_i \partial x_j}+ 
    \frac{\partial ( f_{ij} x_j P_{\rm s})}{\partial 
      x_i}\right)\right], 
\label{eq:a:stationary} 
\eea 
with the boundary condition $ 2 P_{\rm s}(Z_f,x)=P_{\rm s}(Z_i,x)$.  

The instantaneous 
growth rate is given by: 
\beq 
\lambda(x)=\lambda_0+\sum_{i} a_i x_i + \sum_{ij}b_{ij}x_i x_j. 
\eeq 

For the stationary distribution we make the Ansatz  
\beq 
P_{\rm s}(Z,x)\sim 2^{\frac{Z-Z_i}{Z_i-Z_f}}\rme^{-\frac{1}{2}\sum_{ij}\alpha_{ij}(x_i-x^0_i)(x_j-x^0_j)}\sim \rme^{(Z-Z_i)\frac{\log{2}}{Z_i-Z_f}}\rme^{-\frac{1}{2}\sum_{ij}\alpha_{ij}(x_i-x^0_i)(x_j-x^0_j)}. 
\label{eq:a:ans} 
\eeq 
Using the scaling $Z_{\rmi}-Z_{\rm f}= \log(2)$ we obtain 
\beq 
P_s(Z,x) \sim \rme^{-(Z-Z_i)}\rme^{-\frac{1}{2}\sum_{ij}\alpha_{ij}(x_i-x^0_i)(x_j-x^0_j)}. 
\label{eq:a:ans1} 
\eeq 
If we insert this Ansatz into Equation (\ref{eq:a:stationary}), we obtain  
\beq 
g=\lambda_0+\sum_{i} a_i x_i + \sum_{ij}b_{ij}x_i x_j-\sum_{ij}D_{ij}\alpha_{ij}+\sum_{i}f_{ii}+ 
\sum_{ijkl}D_{ij}\alpha_{ik}(x_k-x^0_k)\alpha_{jl}(x_l-x^0_l)-\sum_{ijk}f_{ij} x_j \alpha_{ik}(x_k-x^0_k). 
\eeq 
\end{widetext} 
For this multidimensional polynomial equation to be satisfied for all the values of $\bol{x}$ we must have that all the coefficients are zero. Therefore 
 the growth rate is given by: 
\beq 
g=\lambda_0 -\sum_{ij}D_{ij} \alpha_{ij} +\sum_if_{ii}+\sum_{ijkl}D_{ij}\left(\alpha_{ik}x^0_k\right)\left(\alpha_{jl}x^0_l\right), 
\label{eq:growth} 
\eeq 
where the constants $\bol \alpha$ and ${\bol x}^0$ are obtained by solving 
the set of $\frac{n(n+3)}{2}$ equations: 
\bea 
0&=&a_i-2\sum_{jkl}D_{jk}\alpha_{jl}x_l^0\alpha_{ki}+\sum_{jk}f_{ji}\alpha_{jk}x_k^0, \, \, \forall i, \nonumber\\ 
0&=&b_{ij}+\sum_{kl}D_{kl}\alpha_{ki}\alpha_{lj}-\sum_{k}f_{ki}\alpha_{kj}, \quad \forall i,j. 
\label{eq:tosolve} 
\eea  
We can read from the Equations (\ref{eq:tosolve}) that negative 
curvatures of the instantaneous advancement rate ($b_i<0$) concentrate 
the Gaussian stationary distribution $P_s(Z,x)$ (induce larger 
$\alpha$'s), while non-zero values for $a_i$ displace the averages 
$x_i^0$ of the 
Gaussian stationary distribution $P_s(Z,x)$ such that $a_i x_i^0 >0$. 

\noindent{\bf Growth rate controlled by a single enzyme}\\ 
We derive here Equation (\ref{eq:simpleq}). As discussed in the text,  we model  
the dynamics of enzyme X via the linearized Langevin dynamics,
\begin{equation} 
\dot x = -\gamma x + \eta, 
\end{equation} 
while we assume that  the growth 
rate of a single cell as a function of the expression level of X can be 
written as 
 \beq \lambda=\lambda_0 (X_{\rm s}) + a x + b x^2. 
\eeq 
We must solve the equation 
\bea 0&=& -\frac{\partial (\lambda 
  P_{\rm s})}{\partial Z} - g P_{\rm s}+\left[\left( 
    D \frac{\partial^2 P_{\rm s}}{ \partial x^2}+ 
    \frac{\partial ( \gamma x P_{\rm s})}{\partial 
      x}\right)\right], 
\label{eq:stationary1} 
\eea where we choose $D$ such that the strength of the biochemical 
noise $\eta$ is $2 D$ \cite{Gillespie00}. To obtain the stationary 
distribution, we make the Ansatz \beq P_{\rm 
  s}(Z,x)\sim \rme^{-(Z-Z_i)}\rme^{-\frac{1}{2 \sigma^2} (x-x^0)^2}. \eeq If 
we insert this into Equation (\ref{eq:stationary1}), we find that we 
have to solve the equations \bea 
g&=&\lambda_0 -D/\sigma^2  + \gamma + D \left(\frac{x^0}{\sigma^2}\right)^2,\nonumber\\ 
0&=&a-2D x^0/\sigma^4+ \gamma x^0/\sigma^2, \nonumber\\ 
0&=&b+D/\sigma^4-\gamma/\sigma^2, \eea from which we obtain the solution 

\bea 
g&=&\lambda_0 (X_{\rm s}) +\frac{a^2 D}{\gamma^2-4 b D}+b \sigma^2,\\ 
\sigma^2&=&\frac{2D}{\gamma+\sqrt{\gamma^2-4 b D}}. \eea 
\noindent {\bf Cost-benefit analysis of gene regulation}\\ 
We now present the derivation and the approximations 
leading to Equation (\ref{eq:final}).  A mean-field analysis of the 
cost-benefit function of Dekel and Alon \cite{Dekel05b}, 
Equation (\ref{eq:CB}), predicts that the maximum growth rate occurs at 
$E_{\rm opt}^0=M\left(1-\sqrt{\frac{\eta}{\delta}}\right)$ and 
$X=0$. We are interested in the growth rate of a cell in which the 
average enzyme concentration is $E_{\rm s} = E_{\rm opt}^0$, while the 
average transcription factor concentration, $X_{\rm s}$, is not zero, 
but finite. Since the average transcription factor concentration, 
$X_{\rm s}$, is nevertheless small, it is reasonable to assume that 
the growth rate of a cell with $E=E_{\rm s} + e$ and $X=X_{\rm s}+{\rm 
  x}$ can be obtained by Taylor expanding the growth rate given by 
Equation (\ref{eq:CB}) around the deterministic prediction, $E=E_{\rm s} = 
E_{\rm opt}^0$, $X=0$. This yields \beq 
\lambda=\lambda_{\rm D}+\lambda_0\left[ a_1 e + a_2 x + b (x+e)^2\right], 
\label{expansion} 
\eeq 
where 
\beq 
a_1=0, \quad a_2=-\delta, \quad b=-\frac{\delta}{M} 
\sqrt{\frac{\delta}{\eta}}. 
\label{eq:coefficients} 
\eeq Here, $\lambda_0$ is the growth rate of each single cell when the 
gene regulatory protein and the enzyme are not expressed 
\cite{Dekel05}. The rate $\lambda_{\rm D}$ is the ``deterministic'' 
growth rate, thus the growth rate when the 
regulatory protein and the enzyme are expressed, but fluctuations are 
not taken into account. It is given by: \beq 
\lambda_{\rm D} =\lambda_0+\lambda_0\left[(\sqrt{\delta}-\sqrt{\eta})^2 M 
  -\delta X_{\rm s}- \frac{\delta}{M} \sqrt{\frac{\delta}{\eta}} 
  X_{\rm s}^2\right]. 
\label{eq:lambda_o} 
\eeq 
Remark that at zero $X_{\rm s}$  we have: 
\beq 
\frac{\lambda_{\rm D}-\lambda_0}{\lambda_0}=(\sqrt{\delta}-\sqrt{\eta})^2 M. 
\eeq 

Equations (\ref{eq:growth}) and (\ref{eq:tosolve}) can now be solved using 
Equations (\ref{expansion}--\ref{eq:lambda_o}) to obtain the 
growth rate that takes into account the noise.  This leads to the 
following expression for the growth rate: \bea \frac{ 
  g-\lambda_0}{\lambda_0}=M(\sqrt{\delta}-\sqrt{\eta})^2 - 
\delta X_{\rm s} - \frac{\delta}{M} \sqrt{\frac{\delta}{\eta}} X_{\rm 
  s}^2-\frac{\delta}{2 M} \sqrt{\frac{\delta}{\eta}}\left[\frac{f^2}{\gamma^2}+ 
  \frac{2 f}{\gamma}+2\right]\sigma_{\rm X}^2 +\frac{\lambda_0}{\gamma}\delta^2 
\sigma_{\rm X}^2. 
\label{app:final} 
\eea In deriving Equation (\ref{app:final}) we also use the fact 
that the transcription factor concentration is much smaller than the 
typical enzyme concentration, yielding $\frac{X_{\rm s}}{M}\ll 1$. We 
also use the inequalities $\delta,\eta < \frac{1}{M}$ \cite{Dekel05} 
and the Poissonian nature of the noise in the transcription factor: 
$\sigma_{\rm X}^2=\frac{N_{\rm X}}{V^2}$. Equation (\ref{app:final}) can 
be further simplified by keeping in our approximation only the terms 
of order one or larger in the small ratio $\frac{X_{\rm 
    s}}{M}$. Please note that in the absence of fluctuations, the 
above equation reduces to $g = \lambda_{\rm D}$: the growth rate of the 
population of cells, $g$, then equals the growth rate of each single 
cell, $\lambda_{\rm D}$.

The last term in Equation (\ref{app:final}) is positive, and, 
interestingly, {\em promotes} fluctuations in $X$. It comes from the 
finite derivative at $X=0$, as explained in {\em Biochemical noise 
can  both reduce and enhance the population's growth rate}. However,  \beq 
\sigma_{\rm X}^2\delta^2 <\frac{\sigma_{\rm X}^2}{M^2} =\frac{\sigma_{\rm X}^2}{X_{\rm 
    s}^2}\frac{X_{\rm s}^2}{M^2} \simeq \frac{1}{N_{\rm X}} \frac{X_{\rm 
    s}^2}{M^2}.  \eeq Therefore, the last term in Equation (\ref{app:final}) is negligible at our level of 
approximation. 

We also have  
\beq 
2 \frac{\delta}{M} \sqrt{\frac{\delta}{\eta}} X_{\rm 
  s}^2<2  \sqrt{\frac{\delta}{\eta}}\frac{X_{\rm 
  s}^2}{M^2} \simeq 2 \frac{X_{\rm 
  s}^2}{M^2}, 
\eeq 
while 
\beq 
2 \frac{\delta}{M} \sqrt{\frac{\delta}{\eta}} \sigma_{\rm X}^2<2  \frac{X_{\rm 
  s}^2}{M^2}  \frac{1}{N_{\rm X}}. 
\eeq 
We can therefore simplify Equation (\ref{app:final}) to the form 
\beq 
\frac{g-\lambda_0}{\lambda_0}=M(\sqrt{\delta}-\sqrt{\eta})^2 - 
\delta X_{\rm s}  
-\frac{\delta}{2 M} \sqrt{\frac{\delta}{\eta}}\left[\frac{f^2}{\gamma^2}+ 
 \frac{2 f}{\gamma}\right]\sigma_{\rm X}^2. 
\label{app:final2} 
\eeq 

Around the steady state $\frac{f}{\gamma} \equiv \frac{\partial 
  E}{\partial X_{\rm s}}$, and we thus also have 
$\frac{f^2}{\gamma^2}\gg\frac{f}{\gamma}\simeq \frac{M}{X_{\rm s}}\gg 
1$, such that we can simplify Equation (\ref{app:final2}) further. 
Nevertheless, it is important to remark that positive regulation 
($f>0$) increases the detrimental effect of fluctuations in the 
concentration of the gene regulatory protein. Hence,  
at this level the cost of biochemical noise is smaller for repressors 
than for activators. Finally, Equation (\ref{eq:final}) of the main text is 
obtained by neglecting the term $\frac{f}{\gamma}$ in Equation 
(\ref{app:final2}). 

If the response times of the enzyme and the transcription factor are 
not equal, the same analysis gives \beq 
\frac{g-\lambda_0}{\lambda_0}=M(\sqrt{\delta}-\sqrt{\eta})^2 - \delta 
X_{\rm s} -\frac{\delta}{ M(1+\frac{\gamma_{\rm x}}{\gamma_{\rm 
      E}})}\sqrt{\frac{\delta}{\eta}}\left[\frac{f^2}{\gamma_{\rm 
      E}^2}+ \frac{2 f}{\gamma_{\rm E}}\right]\sigma_{\rm X}^2, \eeq 
where $\gamma_{\rm X}$ is the degradation rate, i.e. the response 
time, of the transcription factor and $\gamma_{\rm E}$ is the 
degradation rate (response time) of the enzyme. This shows that the 
effect of the fluctuations in the transcription factor concentration, 
$X$, critically depends upon the response times of $X$ and $E$: only when $X$ 
fluctuates more slowly than the time scale on which $E$ can respond to 
these fluctuations ($\gamma_{\rm X} < \gamma_{\rm E}$), 
are the fluctuations in $X$ propagated effectively to fluctuations 
in $E$. In contrast, if the fluctuations in $X$ are fast compared to 
the response time of $E$ ($\gamma_{\rm X} > \gamma_{\rm E}$), then the slow 
enzyme dynamics will effectively integrate out the fluctuations in 
$X$; indeed, the last term on the right hand side of the above equation is then small. \\[0.3cm]

\noindent{\bf Acknowledgments}\\ 
We thank Daan Frenkel and Frank Poelwijk for a critical reading of the 
manuscript. This work is part of the research 
program of the ``Stichting voor Fundamenteel Onderzoek der Materie 
(FOM)", which is financially supported by the ``Nederlandse 
organisatie voor Wetenschappelijk Onderzoek (NWO)''.\\

\newpage 

\begin{center} 
\textbf{LIST OF FIGURES} 
\end{center} 

\begin{enumerate} 
\item 
A sketch of the instantaneous growth rate $\lambda$ of a 
  single cell as a function of the concentration $X$ of component 
  X. If the average expression level $X_{\rm s}$ is close to the 
  optimal expression level $X_{\rm opt}$, biochemical noise will 
  always decrease the growth rate. If, however, the average expression 
  level deviates sufficiently from the optimal expression level 
  (i.e. if $ax > bx^2$ in Equation \ref{eq:lambda_X}), then fluctuations 
  can enhance the growth rate of the population, even when the growth 
  rate $\lambda$ of a single cell is linear in $X$, i.e. if 
  $b=0$. The reason is that fast growing cells dominate the 
  population. 
\label{fig:sketch} 

\item 
Relative change in the growth rate as a function of the 
  average repressor concentration. The growth rate is 
  averaged over different lactose concentrations in the 
  environment (see Equation \ref{eq:avg}), for two different lactose 
  concentration distributions in the environment. 
\label{fig:unifvsnonunif} 

\item 
Different regulatory networks can yield the same optimal 
  enzyme expression level as a function of inducer concentration. This 
  is illustrated for two regulatory networks of the \emph{lac} system, which 
  differ in the dissociation constants of lactose-repressor binding 
  and repressor-operator binding. Panels a) and b) show the response 
  functions at two different stages of the \emph{lac} regulatory network, 
  while panel c) shows the resulting optimal enzyme expression level 
  as a function of lactose concentration.  a) The fraction of 
  repressor that is not bound by lactose, $X_{\rm free}/X$, as a 
  function of lactose concentration for two different lactose-repressor 
  binding constants. b) The corresponding response curves of the 
  enzyme expression level as a function of the fraction of free 
  repressor. The total expression level of repressor is chosen to 
  correspond to the optimal growth rate (see Figure 
  \ref{fig:unifvsnonunif}). c) The resulting optimal enzyme expression 
  level as a function of the lactose concentration, as predicted by 
  Equation (\ref{eq:Eopt}) \cite{Dekel05b}.  
\label{fig:Regulation} 

\item 
The optimal design of the \emph{lac} regulatory network is 
  determined by the \emph{lac} repressor copy number and the 
  repressor-lactose binding constant. Contour plot of the growth rate 
  as a function of the repressor copy number $X$ and repressor-lactose 
  binding constant $K_{\rm D}$. The weighting of the lactose levels is 
  nonuniform. Lower binding constants allow for higher optimal growth 
  rates at lower optimal expression levels for the repressor. 
\label{fig:contourplot}

\end{enumerate} 

\newpage 
\vspace*{2cm} 
\begin{figure}[h] 
\includegraphics[width=12cm]{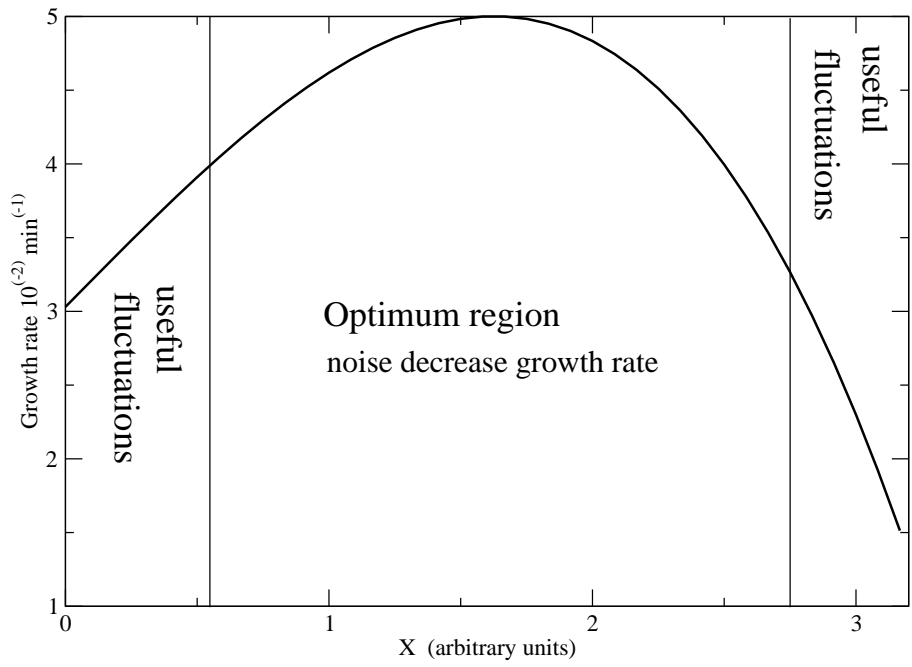} 
\caption{T\u{a}nase-Nicola and Ten Wolde} 
\end{figure} 

\clearpage 

\newpage 

\vspace*{2cm} 
\begin{figure}[h] 
\includegraphics[width=12cm,angle=-90]{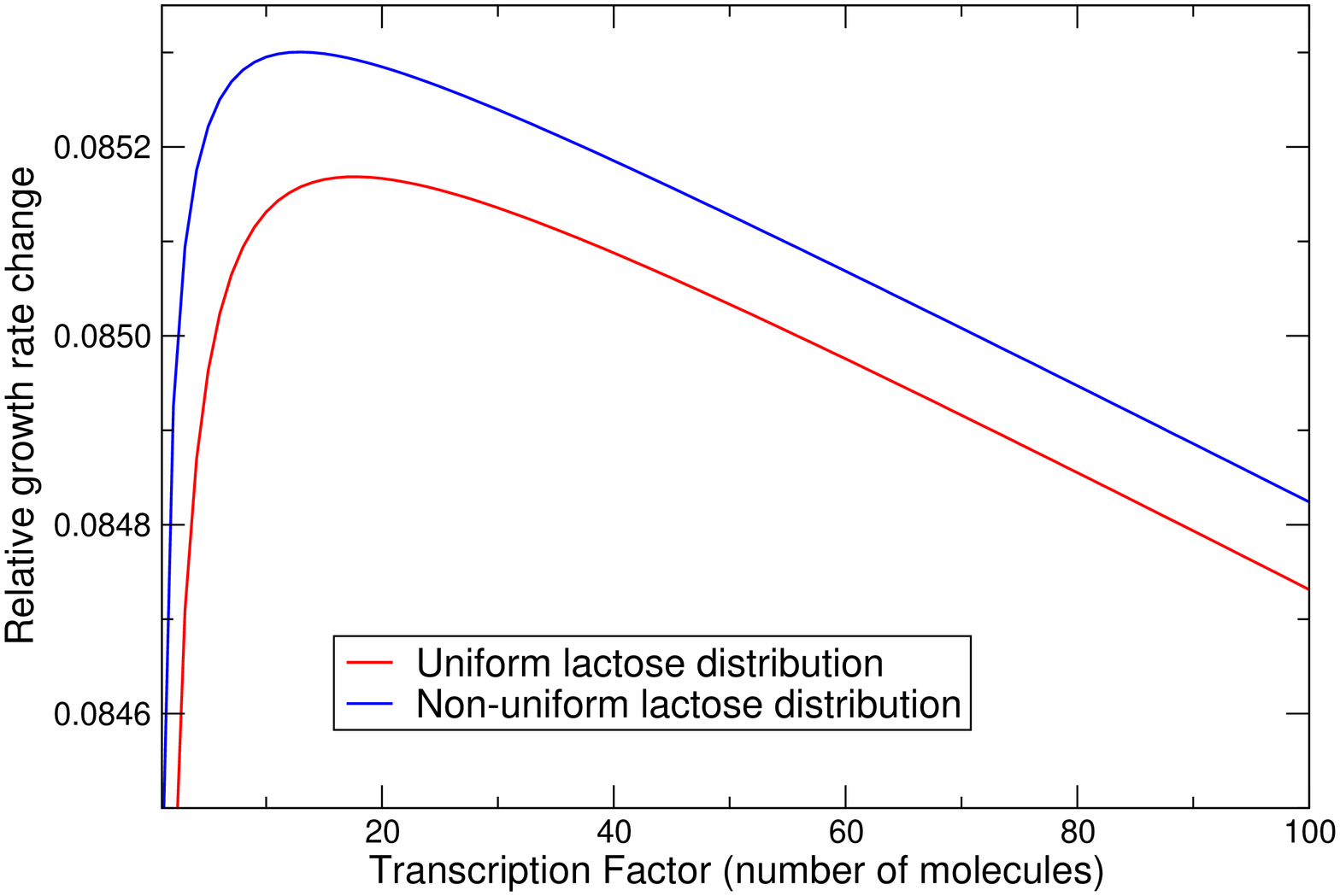} 
\caption{T\u{a}nase-Nicola and Ten Wolde} 
\end{figure} 

\clearpage 

\newpage 

\vspace*{2cm} 
\begin{figure}[h] 
\begin{center} 
\includegraphics[width=12cm]{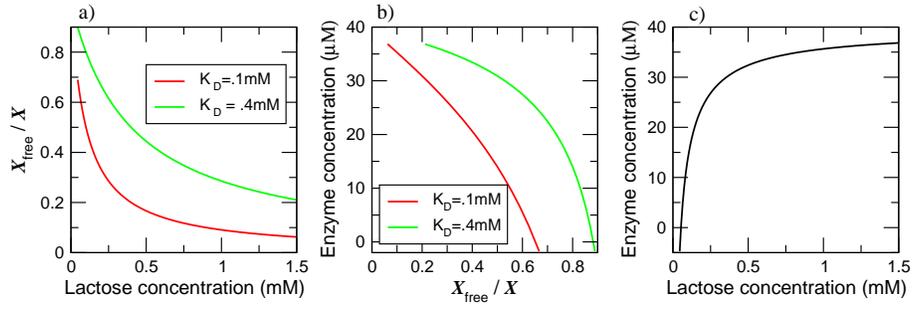} 
\caption{T\u{a}nase-Nicola and Ten Wolde} 
\end{center} 
\end{figure} 

\newpage 
\clearpage 

\vspace*{2cm} 
\begin{figure}[h] 
\includegraphics[width=12cm]{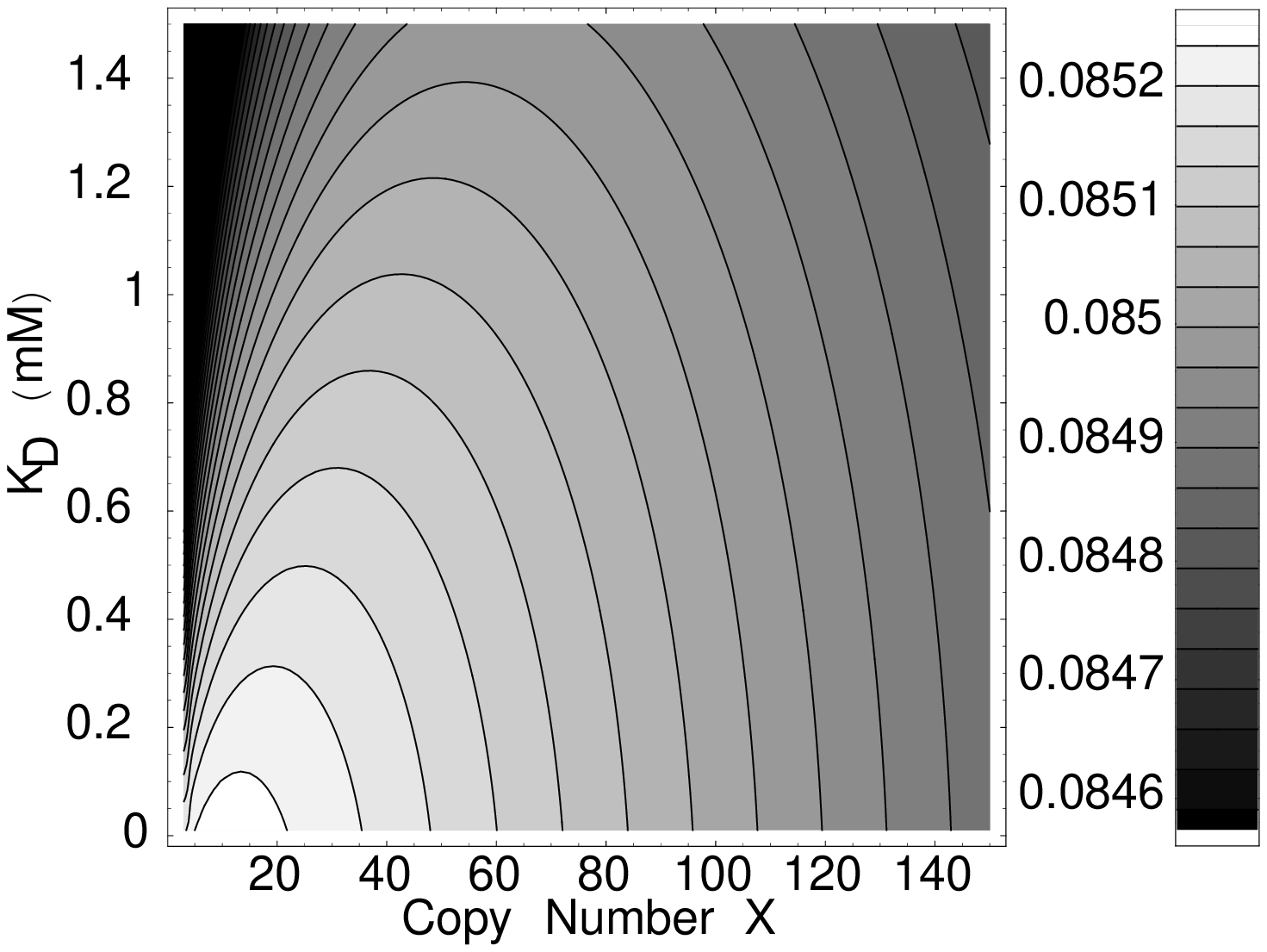} 
\caption{T\u{a}nase-Nicola and Ten Wolde} 
\end{figure}


\begin{thebibliography}{10}
\providecommand{\url}[1]{\texttt{#1}}
\providecommand{\urlprefix}{URL }
\expandafter\ifx\csname urlstyle\endcsname\relax
  \providecommand{\doi}[1]{doi:\discretionary{}{}{}#1}\else
  \providecommand{\doi}{doi:\discretionary{}{}{}\begingroup
  \urlstyle{rm}\Url}\fi
\providecommand{\bibAnnoteFile}[1]{%
  \IfFileExists{#1}{\begin{quotation}\noindent\textsc{Key:} #1\\
  \textsc{Annotation:}\ \input{#1}\end{quotation}}{}}
\providecommand{\bibAnnote}[2]{%
  \begin{quotation}\noindent\textsc{Key:} #1\\
  \textsc{Annotation:}\ #2\end{quotation}}
\providecommand{\eprint}[2][]{\url{#2}}

\bibitem{Kussell05b}
Kussell E, Kishony R, Balaban NQ, Leibler S (2005) Bacterial persistence: A
  model of survival in changing environments.
\newblock Genetics 169:1807--1814.

\input{Kussell05b}

\bibitem{Elowitz00}
Elowitz MB, Leibler S (2000) A synthetic oscillatory network of transcriptional
  regulators.
\newblock Nature 403:335--338.
\input{Elowitz00}

\bibitem{Ozbudak02}
Ozbudak EM, Thattai M, Kurtser I, Grossman AD, van Oudenaarden A (2002)
  Regulation of noise in the expression of a single gene.
\newblock Nat Genet 31:69 -- 73.
\input{Ozbudak02}

\bibitem{Elowitz02}
Elowitz MB, Levine AJ, Siggia ED, Swain PS (2002) Stochastic gene expression in
  a single cell.
\newblock Science 297:1183 -- 1186.
\input{Elowitz02}

\bibitem{Raser05}
Raser JM, O'Shea EK (2005) Noise in gene expression: Origins, consequences, and
  control.
\newblock Science 309:2010--2013.
\input{Raser05}

\bibitem{Kaern05}
Kaern M, Elston TC, Blake WJ, Collins JJ (2005) Stochasticity in gene
  expression: from theories to phenotypes.
\newblock Nat Rev Gen 6:451 -- 464.
\input{Kaern05}

\bibitem{Becskei05}
Becskei A, Kaufmann BB, van Oudenaarden A (2005) Contributions of low molecule
  number and chromosomal positioning to stochastic gene expression.
\newblock Nat Genet 37:937--944.
\input{Becskei05}

\bibitem{Rosenfeld05}
Rosenfeld N, Young JW, Alon U, Swain PS, Elowitz MB (2005) Gene regulation at
  the single-cell level.
\newblock Science 307:1962--1965.
\input{Rosenfeld05}

\bibitem{Yu06}
Yu J, Xiao J, Ren X, Lao K, Xie XS (2006) Probing gene expression in live
  cells, one protein molecule at a time.
\newblock Science 311:1600--1603.
\input{Yu06}

\bibitem{Cai06}
Cai L, Friedman N, Xie XS (2006) Stochastic protein expression in individual
  cells at the single molecule level.
\newblock Nature 440:358--362.
\input{Cai06}

\bibitem{Sigal:2006uq}
Sigal A, Milo R, Cohen A, Geva-Zatorsky N, Klein Y, et~al. (2006) Variability
  and memory of protein levels in human cells.
\newblock Nature 444:643--646.
\input{Sigal:2006uq}

\bibitem{Dekel05b}
Dekel E, Alon U (2005) Optimality and evolutionary tuning of the expression
  level of a protein.
\newblock Nature 436:588--592.
\input{Dekel05b}

\bibitem{Zhang05}
Zhang X, Hill W (2005) Evolution of the environmental component of the
  phenotypic variance: stabilizing selection in changing environments and the
  cost of homogeneity.
\newblock Evolution Int J Org Evolution 59:1237--44.
\input{Zhang05}

\bibitem{Kalisky:2007fj}
Kalisky T, Dekel E, Alon U (2007) Cost-benefit theory and optimal design of
  gene regulation functions.
\newblock Phys Biol 4:229--245.
\input{Kalisky:2007fj}

\bibitem{Falconer1996}
Falconer DS, Mackay TFC (1996) Introduction to quantitative genetics.
\newblock Longman, Harlow, U.K.
\input{Falconer1996}

\bibitem{Gavrilets94}
Gavrilets S, Hastings A (1994) A quantitative-genetic model for selection on
  developemental noise.
\newblock Evolution 48:1478--1486.
\input{Gavrilets94}

\bibitem{Segerbook}
Seger J, Brockman HJ (1987) What is bet-hedging?
\newblock In: Harvey PH, Partridge L, editors, Oxford Surveys in Evolutionary
  Biology. Oxford Univ. Press, Oxford, volume~4, pp. 182--211.
\input{Segerbook}

\bibitem{Thattai04}
Thattai M, van Oudenaarden A (2004) Stochastic gene expression in fluctuating
  environments.
\newblock Genetics 167:523--530.
\input{Thattai04}

\bibitem{Wolf05}
Wolf DM, Vazirani VV, Arkin AP (2005) Diversity in times of adversity:
  probabilistic strategies in microbial survival games.
\newblock Journal of Theoretical Biology 234:227--253.
\input{Wolf05}

\bibitem{Kussell05a}
Kussell E, Leibler S (2005) Phenotypic diversity, population growth, and
  information in fluctuating environments.
\newblock Science 309:2075--2078.
\input{Kussell05a}

\bibitem{vanderWoude04}
van~der Woude MW, Baumler AJ (2004) {Phase and Antigenic Variation in
  Bacteria}.
\newblock Clin Microbiol Rev 17:581--611.
\input{vanderWoude04}

\bibitem{Balaban04}
Balaban NQ, Merrin J, Chait R, Kowalik L, Leibler S (2004) Bacterial
  persistence as a phenotypic switch.
\newblock Science 305:1622--1625.
\input{Balaban04}

\bibitem{Lu07}
Lu T, Shen T, Bennett MR, Wolynes PG, Hasty J (2007) Phenotypic variability of
  growing cellular populations.
\newblock Proc Natl Acad Sci U S A 104:18982--18987.
\input{Lu07}

\bibitem{Gilbert66}
Gilbert W, Muller-Hill B (1966) Isolation of the lac repressor.
\newblock Proc Natl Acad Sci U S A 56:1891--1898.
\input{Gilbert66}

\bibitem{Ingrahambook}
Ingraham JL, Maaloe OL, Neidhardt FC (1993) Growth of the bacterial cell.
\newblock Sinauer Ass., Sunderland, Massachusetts.
\input{Ingrahambook}

\bibitem{Risken96}
Risken H (1996) The Fokker-Plank equation.
\newblock Springer.
\input{Risken96}

\bibitem{Elf03}
Elf J, Ehrenberg M (2003) Fast evaluation of fluctuations in biochemical
  networks with the linear noise approximation.
\newblock Genome Res 13:2473 -- 2482.
\input{Elf03}

\bibitem{Tanase06a}
T\u{a}nase-Nicola S, Warren PB, ten Wolde PR (2006) Signal detection,
  modularity, and the correlation between extrinsic and intrinsic noise in
  biochemical networks.
\newblock Phys Rev Lett 97:068102.
\input{Tanase06a}

\bibitem{Ziv07}
Ziv E, Nemenman I, Wiggins CH (2007) Optimal signal processing in small
  stochastic biochemical networks.
\newblock PLoS ONE 2:e1077.
\input{Ziv07}

\bibitem{Austin06}
Austin DW, Allen MS, McCollum JM, Dar RD, Wilgus JR, et~al. (2006) Gene network
  shaping of inherent noise spectra.
\newblock Nature 439:608--611.
\input{Austin06}

\bibitem{Bar-Even06}
Bar-Even A, Paulsson J, Maheshri N, Carmi M, O'Shea E, et~al. (2006) Noise in
  protein expression scales with natural protein abundance.
\newblock Nat Genet 38:636--643.
\input{Bar-Even06}

\bibitem{Yagil71}
Yagil G, Yagil E (1971) On the relation between effector concentration and the
  rate of induced enzyme synthesis.
\newblock Biophys J 11:11--27.
\input{Yagil71}

\bibitem{Newman06}
Newman JRS, Ghaemmaghami S, Ihmels J, Breslow DK, Noble M, et~al. (2006)
  Single-cell proteomic analysis of s. cerevisiae reveals the architecture of
  biological noise.
\newblock Nature 441:840--846.
\input{Newman06}

\bibitem{Wagner05}
Wagner A (2005) Energy constraints on the evolution of gene expression.
\newblock Mol Biol Evol 22:1365--1374.
\input{Wagner05}

\bibitem{Savageau74}
Savageau MA (1974) Genetic regulatory mechanisms and the ecological niche of
  esche richia coli.
\newblock Proc Natl Acad Sci U S A 71:2453--2455.
\input{Savageau74}

\bibitem{Savageau77}
Savageau MA (1977) Design of molecular control mechanisms and the demand for
  gene expression.
\newblock Proc Natl Acad Sci U S A 74:5647--5651.
\input{Savageau77}

\bibitem{Detwiler00}
Detwiler PB, Ramanathan S, Sengupta A, Shraiman BI (2000) Engineering aspects
  of enzymatic signal transduction: Photoreceptors in the retina.
\newblock Biophys J 79:2801--2817.
\input{Detwiler00}

\bibitem{Paulsson04}
Paulsson J (2004) Summing up the noise in gene networks.
\newblock Nature 427:415 -- 418.
\input{Paulsson04}

\bibitem{Pedraza05}
Pedraza JM, van Oudenaarden A (2005) Noise propagation in gene networks.
\newblock Science 307:1965--1969.
\input{Pedraza05}

\bibitem{Shibata05}
Shibata T, Koichi F (2005) Noisy signal amplification in ultrasensitive signal
  transduction.
\newblock Proc Natl Acad Sci U S A 102:331 -- 336.
\input{Shibata05}

\bibitem{Levine07}
Levine E, Hwa T (2007) Stochastic fluctuations in metabolic pathways.
\newblock Proc Natl Acad Sci U S A
  104:9224 -- 9229 

\input{Levine07}

\bibitem{Kolkhof92}
Kolkhof P (1992) Specificities of three tight-binding lac repressors.
\newblock Nucl Acids Res 20:5035--5039.
\input{Kolkhof92}

\bibitem{Sadler83}
Sadler JR, Sasmor H, Betz JL (1983) A perfectly symmetric lac operator binds
  the lac repressor very tightly.
\newblock Proc Natl Acad Sci U S A 80:6785--6789.
\input{Sadler83}

\bibitem{Gillespie00}
Gillespie DT (2000) The chemical langevin equation.
\newblock J Chem Phys 113:297 -- 306.
\input{Gillespie00}

\bibitem{Dekel05}
Dekel E, Mangan S, Alon U (2005) Environmental selection of the feed-forward
  loop circuit in gene-regulation networks.
\newblock Physical Biology 2:81--88.
\input{Dekel05}

\end{thebibliography}
\end{document}